\documentclass[useAMS,usenatbib]{mn2e}
\usepackage{graphicx,amssymb}

\usepackage{times}

\newcommand{\sm}{\, {\rm M}_{\odot}}

\newcommand{\minnerdm}{M_\mathrm{{0.6}}}

\newcommand{\kms}{km~s$^{-1}$}

\title[On the common mass scale of the MW satellites]{On the common mass scale of
  the Milky Way satellites} 
\author[Y.-S. Li et al.]
       {Yang-Shyang Li$^{1}$\thanks{E-mail: ysleigh@astro.rug.nl (Y-SL); ahelmi@astro.rug.nl (AH)}, 
        Amina Helmi$^{1}$\footnotemark[1],
        Gabriella De Lucia$^{2}$\thanks{Present address: INAF - Astronomical Observatory of Trieste, via
G.B.\ Tiepolo 11, I-34143 Trieste, Italy.} and 
	Felix Stoehr$^{3}$\\
        $^{1}$Kapteyn Astronomical Institute, University of Groningen, P.O.Box
        800, 9700 AV Groningen, the Netherlands\\
        $^{2}$Max--Planck--Institut f\"ur Astrophysik, Karl--Schwarzschild--Str. 1, D-85748 Garching, Germany\\
	$^{3}$European Organization for Astronomical Research in the Southern Hemisphere, 
	Karl--Schwarzschild--Str. 2,
	D-85748 Garching, Germany
}
\begin{document}

\pagerange{\pageref{firstpage}--\pageref{lastpage}} \pubyear{2009}

\maketitle

\label{firstpage}

\begin{abstract}
We use a hybrid approach that combines high-resolution simulations of
the formation of a Milky Way-like halo with a semi-analytic model of
galaxy formation to study the mass content of dwarf galaxies in the
concordance $\Lambda$CDM cosmology. We find that the mass within
600~pc of dark matter haloes hosting luminous satellites has a median
value of $\sim 3.2\times 10^7 \sm$ with very little object-to-object
scatter. In contrast, the present day total luminosities of the model
satellites span nearly five orders of magnitude. These findings are in
very good agreement with the results recently reported in the
literature for the dwarf spheroidal galaxies of the Milky Way. 
In our model, dwarf irregular galaxies 
like the Small Magellanic Cloud, are predicted to have similar or slightly
larger dark matter mass within 600~pc.
\end{abstract}


\section{Introduction}

The dwarf satellite galaxies of the Milky Way are the most dark matter
dominated systems known to date in the Universe. They represent a
heterogeneous population in terms of their stellar properties such as
luminosity, star formation and chemical enrichment histories
\citep{mateo,dolphin,martin}. Yet, the mass enclosed within a radius of 300 (or
600)~pc appears to be roughly constant
\citep{gilmore07,strigari07,strigari08}. This could imply a minimum mass scale
for the existence of dwarf spheroidal galaxies, as originally suggested by
\citet{mateo}. It is currently unclear whether this is due to the microphysics
of the dark matter particles, or to astrophysical processes that inhibit star
formation on small scales. 

For Weakly Interacting Massive Particles (WIMPs), 
collisional damping and free streaming are expected to cut off the power spectrum at masses 
of $\sim 10^{-6} \sm$ \citep[e.g.][]{green} or smaller.  Then, in models 
which assume these as dark matter particles [e.g.\ warm dark matter (WDM), $\Lambda$ cold dark matter (CDM)], 
a minimum mass scale for dwarf spheroidals can only result
as a consequence of astrophysical processes that affect the collapse
of baryons and the formation of stars on small galactic scales. For
example, the presence of a strong photo-ionizing background (possibly
associated to the reionization of the Universe) can suppress 
accretion and cooling in low-mass haloes. This is because the heating
of the gas will raise its pressure and therefore may suppress its
collapse in haloes with virial velocities $\lesssim 30-50\,{\rm
  km}{\rm s}^{-1}$ 
\citep[e.g.][and references
  therein]{Efstathiou_1992,Quinn_Katz_Efstathiou_1996,Thoul_Weinberg_1996,Gnedin_2000,Okamoto_Gao_Theuns_2008}. 
This
has often been considered as a possible solution to the excess of
small scale structures found in CDM and in particular in $N$-body
simulations of galaxy size systems
\citep{kwg,klypin99,moore99,bullock00,somerville,benson}. 
In addition, in systems with virial temperature below $10^4$~K gas cannot
cool via hydrogen line emission, and must rely on the highly
inefficient cooling through collisional excitations of $\mathrm{H}_2$ molecules
\citep{haiman,kravtsov2004}.

In this Letter we discuss the existence of a common mass scale for
Milky Way satellites using results from high resolution $N$-body
simulations of galaxy-size haloes coupled with semi-analytic
techniques to model the evolution of the baryonic component of
galaxies. This approach allows us to identify the dark matter
substructures that host stars and to characterise their stellar
properties. At the same time, the high resolution of the simulations
used in this study permits a reliable determination of the internal
dynamical properties of these satellites. As we shall describe below,
we find that the dark matter mass within 600~pc for the model
satellites shows very little scatter from object to object. This is in
very good agreement with the observational results by
\citet{strigari07,strigari08}. Interestingly, our model also
reproduces the very wide range of luminosities observed for the
satellite galaxies around the Milky Way.

\section{The Simulation and the Galaxy Formation Model}
\label{sec:simsam}

In this study, we use a high-resolution resimulation of a `Milky Way'
halo from the GA series described in \citet{Stoehr_etal_2002} and
\citet{Stoehr_etal_2003}. The candidate `Milky Way' halo was selected
as a relatively isolated halo with a `quiet' merging history (last
major merger at $z > 2$) and with maximum rotational velocity close to
$\sim 220$ \kms. The halo, selected from an intermediate resolution
cosmological simulation, was then re-simulated at four progressively
higher resolutions using the `zoom' technique
\citep*{Tormen_etal_1997}. The underlying cosmological model is a flat
$\Lambda$-dominated CDM Universe with cosmological parameters:
$\Omega_{\rm m}=0.3$, $\Omega_{\Lambda}=0.7$, $H_0 = 70\,{\rm
  km}\,{\rm s}^{-1}\,{\rm Mpc}^{-1}$, $n=1$, and $\sigma_8=0.9$. In
this study, we use the highest resolution simulation of the series -
GA3new - which contains $\sim 10^7$ particles within the virial
radius.

As explained in \citet{gdl_ah}, the simulated halo is more massive 
($M_{200}(z=0) \sim 3 \times 10^{12} \sm$, - $M_{200}$ is here 
defined as the mass 
of a spherical region with interior average density 200 times the critical 
density of the Universe at redshift $z$) 
than recently estimated for our Galaxy, e.g.\ \cite{Battaglia, Battaglia_erratum} and \cite{Xue2008}.
\footnote{We note that \cite{Battaglia} and \cite{Xue2008} use a different 
definition of virial mass which is 100 times of 
the critical density.  Our halo mass is still larger than observational 
estimates when the same definition is adopted.}  
Following
\citet{Helmi_White_and_Springel_2003}, we then scale our `Milky Way' halo by
adopting a scaling factor in mass ${\rm M}_{200}/ {\rm M}_{\rm MW} = \gamma^3 =
2.86$. This implies that we scale down the positions and velocities by a factor
$\gamma = 1.42$. The Plummer equivalent softening length for the scaled
simulation is $0.18$~kpc. The scaled particle mass is $1.03\times 10^5 \sm$.

Simulation data (stored in 108 outputs between $z=37.6$ and $z=0$) were
analysed using a standard friends-of-friends algorithm and the substructure
finder algorithm {\small SUBFIND} \citep{Springel_etal_2001}. As in previous
work, we have considered all substructures retaining at least 20 self-bound
particles - which sets the substructure mass limit to $2.06\times10^6 \sm$, for
the scaled simulation. Finally, these halo catalogues have been used to
construct the halo merger trees that represent the basic input for the galaxy
formation model used in this study. For details on the post-processing, we
refer the interested reader to \citet{Springel_etal_2005} and to
\citet{DeLucia_Blaizot_2007}. 

The galaxy formation model used in our study is a refinement of the model
described in \citet{gdl_ah}, who have studied the predicted physical properties
of the Milky Way and of its stellar halo using the same set of re-simulations
used here. This model builds upon the methodology introduced in
\citet{Springel_etal_2001} and \citet{DeLucia_Kauffmann_White_2004} and has
been further refined in later years. The interested reader is referred to
\citet{Croton_etal_2006}, \citet{DeLucia_Blaizot_2007}, and \citet{gdl_ah} for
a detailed account of the modelling of the various physical processes
considered. In order to obtain a better agreement with the observed properties
of the Milky Way satellites, a few refinements were made to the model used in
\citet{gdl_ah}. For completeness, we give here a short description of these
refinements that will be described in detail in a forthcoming paper (Li et al.,
in preparation). 

\begin{itemize}
\item As described in \citet{Croton_etal_2006}, reionization was
  modelled using the formulation provided by \citet{kravtsov2004} and
  was previously assumed to start at redshift 8 and to be completed by
  redshift 7. In the study presented here, we assume an `early'
  reionization which starts at redshift 15, and is complete by
  redshift 11.5 \citep{wmap3}.
\item In previous studies, haloes with virial temperature below
  10$^4$~K were allowed to cool at the rates corresponding to
  10$^4$~K. In this study, we completely suppress cooling in these
  low-mass haloes \citep[e.g][]{haiman}.
\item In previous implementations of our galaxy formation model, all
  metals produced by new stars were instantaneously mixed to the cold
  gas \citep{DeLucia_Kauffmann_White_2004,gdl_ah}. Inspired by the
  numerical simulations of \citet{maclow-ferrara}, in this study we
  assume that for galaxies embedded in haloes with virial mass below $5
  \times 10^{10} \sm$, most of the new metals (95 per cent) are
  ejected directly into the hot gas phase.
\end{itemize}

As we will show in Li et al.\ (in preparation), these refinements result in a
satellite population with physical properties that closely resemble those
observed for the Milky Way satellites, while leaving essentially unaltered 
the results discussed in \citet{gdl_ah}.

\section{Results}

\begin{figure}
\includegraphics[width=8cm,clip]{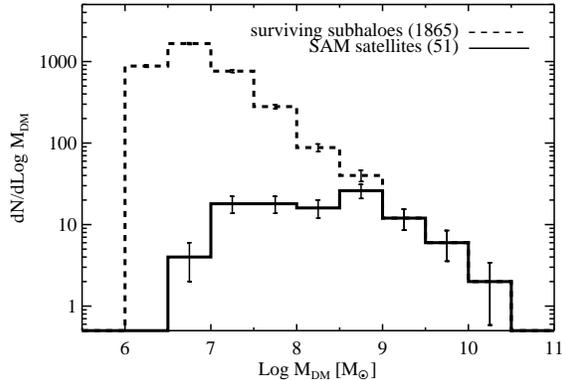}
\caption{The solid histogram shows the present-day mass function for the
  satellites in our model. Error bars show the corresponding 1-$\sigma$ Poisson
  uncertainties. The dashed histogram shows the subhalo mass function, which
  steeply rises up to the resolution limit of our simulation.}
\label{fig:dndm}
\end{figure}

Fig.~\ref{fig:dndm} shows the mass function of all substructures identified at
redshift zero within 280~kpc from the central galaxy and lying in the same FOF
group (dashed histogram), and the corresponding mass function of satellites
(i.e.\ subhaloes hosting stars) in our model (solid histogram). The subhalo
masses plotted in Fig.~\ref{fig:dndm} are computed summing up the masses of all
self-bound particles, and scaled as discussed in Sec.~\ref{sec:simsam}.
Fig.~\ref{fig:dndm} shows that the dark matter masses of our satellites span a
relatively large range from $\sim 5 \times 10^6 \sm$ to $\sim 5 \times 10^{10}
\sm$, and that the distribution is nearly flat between $10^7$ and $10^9
\sm$. In contrast, the subhalo mass function continues to rise steeply up to
the resolution limit of our simulation.  The simulation used in this study
contains almost 2000 subhaloes in the considered region, but our model predicts
that only 51 of them host stars. This is still larger (by a factor of about 2)
than the number of Milky Way satellites currently known. However, when
corrections due to incompleteness are considered, this discrepancy in number
is eliminated.   
For example, \citet{koposov} predict $\sim 45$ satellites down 
to $M_{V}=-5$ magnitude.  This is consistent with the number of satellites 
predicted by \cite{Tollerud_etal} down to the same magnitude limit.  The small
number of faint satellites could be a feature of this particular model, or due
to the numerical resolution of the simulation adopted in our work.  This will 
be tested in future studies using higher resolution simulations. 

The high resolution of the simulation used in this study allows us to
measure the satellites' dark matter mass enclosed within a (small) given
radius. Following \citet{strigari07}, we measure the mass within
600~pc, which corresponds to 3.33 (scaled) softening lengths. We do
not attempt to measure the mass within 300~pc, as done in the more
recent analysis by \citet{strigari08}, as this would be beyond reach
for the GA3new simulation.

In order to measure the mass within 600~pc ($\minnerdm$ hereafter), we
compute the centre of mass position for each subhalo using its 10 per
cent most bound particles. We then count the number of bound particles
located within 600~pc from the centre of mass of each subhalo, and
multiply this number by the particle mass. We find that our subhaloes
have on average $\sim 380$ particles within this distance, and in
nine cases $\lesssim 100$ particles.

Numerical effects will in general tend to artificially lower the mass
in the inner regions of subhaloes.  We believe, however, that our
measurements are robust and are not strongly affected by the limited
numerical resolution. In order to quantify the impact of numerical
resolution on our estimates of $\minnerdm$, we assume that the inner
density profiles of subhaloes are well fit by Einasto profiles, as
recently demonstrated by \citet{springel08} using the very
high-resolution simulations of the Aquarius project. The Einasto
profile can be fully characterised with 3 parameters, namely the
logarithmic slope $\alpha$, the peak circular velocity $V_{\rm max}$
and the radius $r_{\rm max}$ where this peak value is
reached. \citet{springel08} find $\alpha \sim 0.15 - 0.25$, with an
average value of $\sim 0.18$. Assuming this value for $\alpha$, and
measuring $V_{\rm max}$ and $r_{\rm max}$ directly from the simulation
used in this study, we obtain an independent estimate of 
$\minnerdm$.  

When we compare the $\minnerdm$ values derived using the Einasto
profile with those directly measured in the simulation, we find at
most a factor two increase, but in most cases the difference is less
than 40 per cent. The largest deviation is found for a subhalo nearing
disruption and whose circular velocity curve is particularly
noisy. For this object, as well as for those subhaloes with rotation
curves within 600~pc (and $\minnerdm$) that have evolved
significantly in the last 2 Gyrs (these are in general the least
massive objects), we always show both the direct measurement of
$\minnerdm$ and the value derived using the Einasto
profile. For the other satellites, we keep the direct measurements
only. It is important to note that the corrections derived from the
Einasto model are smaller than the scatter found in the
values of $\minnerdm$ measured for model satellites.

\begin{figure}
\includegraphics[width=8cm,clip]{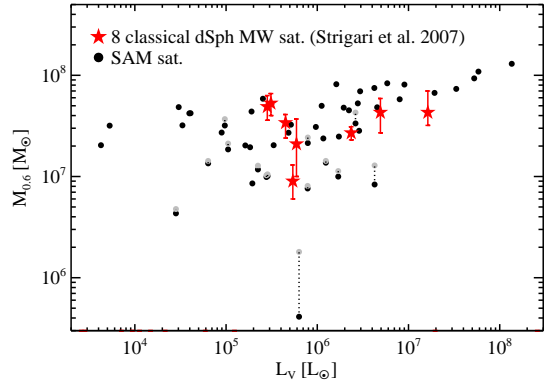}
\caption{Mass within 600~pc as a function of the $V$-band luminosity for
  our model satellites (black and grey solid circles), and for the
  eight brightest dwarf spheroidal galaxies of the Milky Way (asterisks), 
  where the Sagittarius dwarf has been excluded because 
it is clearly not in dynamical equilibrium.  
  Black solid circles correspond to $\minnerdm$ measured via direct particle
  counting.  For the model satellites which show clear signs of tidal
  disruption of the associated subhaloes, we also plot the Einasto
  $\minnerdm$ estimates shown with grey symbols.}
\label{fig:m_600}
\end{figure}

Fig.~\ref{fig:m_600} shows the measured $\minnerdm$ for satellites as
a function of the total $V$-band luminosity predicted by our galaxy
formation model. Black symbols show the values of $\minnerdm$ measured
directly from the simulations. As discussed above, for satellites with
signs of tidal disruption we also show the corresponding values
estimated assuming an Einasto profile with $\alpha = 0.18$. These are
the grey symbols linked to the direct measurement by dotted lines. In
this Figure we have excluded two objects whose stellar masses, as
derived from the galaxy formation model, are larger than the dark
matter mass of the associated subhalo. These objects are nearly fully
disrupted in the numerical simulation, but since we do not follow the
tidal stripping of the stars in our model, we  
could be overestimating 
their current luminosity by an unknown factor.  
However, we find that only $10-15$ satellites show evolution in
$\minnerdm$ and the rotation curves in the last 2 Gyr.  We therefore
suspect that tidal stripping does not affect significantly the
luminosity of model galaxies.

Our model satellites span more than four orders of magnitude in
luminosity, which is comparable to what is observed for the dwarf
galaxies around the Milky Way when one includes the recently
discovered `ultra-faint satellites'. In contrast, their dark matter
masses within 600~pc do not differ by more than one decade. The
asterisks in Fig.~\ref{fig:m_600} show the estimates given by
\citet{strigari07} for the classical dwarf spheroidal satellites of
the Milky Way. Our results are therefore entirely consistent with
previous analyses claiming the existence of a minimum mass scale of
the order of $10^7 \sm$, and no (or very few) subhaloes (in
equilibrium) hosting stars below this threshold
\citep{mateo,strigari07,strigari08,gilmore07}.  We note that the correlation 
between $\minnerdm$ and the $V$-band luminosity in the model is somewhat 
stronger than in the real data.  In the context of our
model, satellites with smaller $\minnerdm$ values could exist, but
these are expected to be strongly tidally disturbed,
i.e.\ out of dynamical equilibrium.  
 
All our satellites were massive enough prior to accretion to be above
the cooling limit set by the atomic hydrogen (e.g.\ $\sim 10^9\sm$ at $z\sim 1$).  
Objects of such mass should have lost more than 99 per cent of their mass in order 
to have $\minnerdm$ smaller than $10^7\sm$ at present time.  
Note that in Fig.~\ref{fig:m_600}
we have also included model satellites that could be the counterparts
of systems like the Small Magellanic Cloud or NGC6822
(i.e.\ gas-rich). In our model, such luminous objects are expected to
be embedded in the most massive subhaloes at the present time. Hence
Fig.~\ref{fig:m_600} shows that we predict that the Small Magellanic
Cloud should have $\minnerdm$ only slightly larger,
namely $\sim 10^8\sm$, than the corresponding values measured for the dwarf spheroidals
surrounding the Milky Way.

\begin{figure}
\includegraphics[width=8cm,clip]{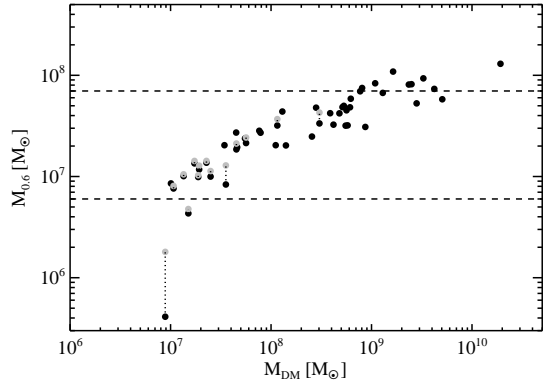}
\caption{Mass within 600~pc as a function of the total mass estimated
  by {\small SUBFIND}. The two dashed lines correspond to the upper
  and lower limits for the observational estimates of $\minnerdm$
  for the 8 classical Milky Way dwarf spheroidal galaxies \citep[dSphs;][]{strigari07}. Symbols are the
  same as in Fig.~\ref{fig:m_600}.}
\label{fig:mvir}
\end{figure}

An intriguing question is whether this result implies that the Milky
Way satellites reside in dark matter haloes of the same total mass. We
address this question in Fig.~\ref{fig:mvir} where we show the
estimated $\minnerdm$ as a function of the total dark matter mass
given by {\small SUBFIND} for the objects shown in
Fig.~\ref{fig:m_600}. While for most model satellites $\minnerdm$
varies in the range $\sim 10^7 - 10^8 \sm$, their  present-day total mass spans
three orders of magnitude (from $\sim 10^7$ to $\sim 10^{10}
\sm$). Therefore, in our model, a common mass within 600~pc does not
imply the same total mass. This is because this scale is generally too
small compared to the full extent of dark matter subhaloes hosting
luminous satellites. In other words, satellites are not embedded in 
dark matter subhaloes that can be characterised by a single 
parameter, since the relationship between concentration and 
virial mass is not very tight for these objects  
before the time of accretion.

Our model satellites are found to be embedded in dark matter haloes
whose present total mass is larger than $\sim 10^7
\sm$. \citet{strigari08} have suggested that this could imply that
these objects had a characteristic mass of $\sim 10^9 \sm$ at the time
of accretion (i.e.\ when they became satellites of the Milky Way). In
our model, we find a broad range of total dark matter mass at the time
of accretion for our luminous satellites, that extends from $\sim 3
\times 10^8\sm$ up to $6 \times 10^{10} \sm$.  Comparison to Fig.~\ref{fig:mvir} indicates that most of the 
satellites have lost (a significant amount of) dark matter due to tidal 
stripping after accretion.  

The existence of a
minimum mass scale for model satellites is essentially the result of a
combination of the two afore mentioned physical processes: cooling is
strongly inhibited for haloes with $T_{\rm vir} < 10^4$~K, and
reionization prevents the further collapse of gas onto low-mass
haloes.  

The wide range of total dark matter masses observed today reflects in
part the initial broad range of masses of the satellites'
haloes. Furthermore, the tidal field of the Galaxy halo will act to
increase the difference in present-day mass of these objects depending
on when they were accreted as well as on their orbits
\citep{DeLucia_etal_2004,Gao_etal_2004}. For example, our faintest
objects, which are typically also the least massive today, were
accreted earlier and have suffered significant stripping. On the other
hand the brightest satellites reside in the most massive haloes, have
typically been accreted only recently and therefore have not been
significantly affected by tides. 

A simple argument that illustrates how the measured range of halo masses at
accretion translates into a wide range of satellite luminosities is as follows.  
We assume that the mass accretion histories of haloes hosting luminous 
satellites can be described as $M(z) = M_0 e^{-2 a_c z}$, prior to accretion 
\citep{Wechsler_etal}. $M_0$ denotes the maximum total mass of a halo that 
is able to grow (without being accreted onto a larger structure) until $z=0$. 
This expression has been shown to be valid for clusters and galaxy-sized 
haloes but we have checked that it is valid also for the haloes hosting 
our satellites, with $a_c \sim 0.2$. We also assume that, at the time of 
accretion, each galaxy's luminosity is proportional to the mass of its 
dark matter halo \citep[see e.g.][]{Wang_etal}.  If we neglect the star 
formation occurring after accretion, we can write $L \propto M_{max}$, 
where $M_{max} = M_0 e^{-2 a_c z_{acc}}$. The range of observed luminosities 
at present day can then be estimated as:
\begin{equation}
L_{\rm low}/L_{\rm up}=[M_{max}^{\rm low}/M_{max}^{\rm
up}]=[M_{0}^{\rm low}/M_{0}^{\rm up}]\times e^{-2a_{c} (\Delta z)}.
\end{equation} 
In this expression, the first term would be related to the range of
total masses of the satellites at fixed accretion epoch, and in the
second term $\Delta z$ denotes the range of accretion redshifts.  
In our simulation, we find $M_{0}^{low}/M_{0}^{up} \sim 70$ and 
$\Delta z \sim 6$, which give a range of luminosities $\sim 10^3$, 
not far from what is observed.

\section{Conclusions}

We have used a high-resolution simulation of the formation of a Milky
Way-like halo in combination with a semi-analytic model of galaxy
formation in order to study the mass content of subhaloes hosting
luminous satellites. The galaxy formation model used in this study
shows considerable agreement with a large number of observed galaxy
properties (see discussion in \citealt{gdl_ah} and references
therein). With the few refinements discussed in Sec~\ref{sec:simsam},
the same model is able to reproduce quite well the physical properties
of the observed Milky Way satellites (Li et al.\ in preparation) while
leaving essentially unaltered the level of agreement with
observational data shown in previous work.

The key result of this Letter is that our model predicts naturally a
common dark matter mass scale within 600~pc for the luminous
satellites.  Our model satellites span nearly five orders of magnitude
in luminosity, while their dark matter masses within 600~pc vary only
by one decade (between $\sim 10^7$ and $\sim 10^8 \sm$), in very good
agreement with recent observational measurements. The total dark
matter masses of our luminous satellites, however, span about three
orders of magnitude with the scatter reflecting the lack of a tight 
concentration-virial mass relation, the different accretion
times, and the different amounts of tidal stripping suffered by the parent
substructures once they have fallen onto the Milky Way halo. The
existence of such a scale in the context of our model results from the
strong suppression of accretion and cooling of gas by low-mass haloes
after reionization, as well as from the atomic hydrogen cooling
threshold at $T_{\rm vir} = 10^4$~K. These physical processes then
inhibit the formation of stars in objects that never reached virial
velocities above $\sim 17$~\kms.

We note that our analysis has been carried out considering the mass
within 600~pc, and an even tighter distribution may be expected when
measuring the mass within 300~pc as used in the most recent study by
\citet{strigari08}. This mass scale is beyond the resolution limit of
the simulation used in this Letter but is within reach of the new
generation of ultra-high resolution simulations like those carried out
within the Aquarius project.

After the submission of this Letter, \cite{mkm09} presented a 
similar study for the satellites of three Milky Way-like haloes that 
reproduces the \cite{strigari08} results.  Although these authors claim 
a relatively narrow distribution of circular velocities at the time of 
accretion, their masses span a similar range as ours.  They also reach a 
similar conclusion as we do.  Namely, that this relation is due to a 
variety of masses at the time of accretion and a broad distribution 
of accretion times.  More recently, \cite{koposov09} has also shown the common mass scale 
for the Milky Way satellites is expected in their CDM based galaxy 
formation models.  It is worthwhile to point out that all these 
models have included the effect of a photoionization background 
and accounted for the inefficient molecular cooling below $10^4$~K.

\section*{Acknowledgments}
Giuseppina Battaglia, Volker Springel and Simon White are
gratefully acknowledged for interesting discussions.  The referee, 
David Weinberg, is warmly thanked for a careful reading 
and suggestions to improve the manuscript.  This work has 
been partially supported by the Netherlands Organisation for Scientific
Research (NWO). AH and GDL wish to thank the hospitality of the KITP, where
this Letter reached its final form.  This research was supported in 
part by the National Science Foundation under Grant No.\ PHY05-51164.

\bibliographystyle{mn2e.bst}
\bibliography{references} 

\label{lastpage}

\end{document}